\documentclass[runningheads]{llncs}
\usepackage{graphicx}
\usepackage{color}
\newif\ifdraft
\drafttrue
\ifdraft
\newcommand{\abhi}[1]{ {\textcolor{red} { ***Abhinav: #1 }}}
\else
\newcommand{\abhi}[1]{ {}}
\fi

\begin{document}
\title{Use Cases for High Performance Research Desktops}
\author{Robert Henschel\inst{1,2} \and
Jonas Lindemann\inst{3} \and Anders Follin\inst{3} \and
Bernd Dammann\inst{4} \and Cicada Dennis\inst{1} \and Abhinav Thota\inst{1}}
\authorrunning{R. Henschel et al.}

%
\institute{Indiana University, Bloomington, IN 47408, USA \email{\{henschel,hbrokaw,athota\}@iu.edu}  \and
Cendio AB, Linköping 58330, Sweden \and
Lund University, Lund 22100, Sweden \email{\{jonas.lindemann,anders.follin\}@lunarc.lu.se}\and
Technical University of Denmark, 2800 Kgs. Lyngby, Denmark \email{beda@dtu.dk} 
}

\maketitle              
\begin{abstract}
High Performance Research Desktops are used by HPC centers and research computing organizations to lower the barrier of entry to HPC systems. These Linux desktops are deployed alongside HPC systems, leveraging the investments in HPC compute and storage infrastructure. By serving as a gateway to HPC systems they provide users with an environment to perform setup and infrastructure tasks related to the actual HPC work. Such tasks can take significant amounts of time, are vital to the successful use of HPC systems, and can benefit from a graphical desktop environment. In addition to serving as a gateway to HPC systems, High Performance Research Desktops are also used to run interactive graphical applications like MATLAB, RStudio or VMD.
This paper defines the concept of High Performance Research Desktops and summarizes use cases from Indiana University, Lund University and Technical University of Denmark, which have implemented and operated such a system for more than 10 years. Based on these use cases, possible future directions are presented.

\end{abstract}
\section{Introduction}
This paper employs use cases to illustrate the features of High Performance Research Desktops and to underscore the real-world impacts of enabling such use cases. The use cases are largely coming from three organizations that have been running a High Performance Research Desktop for more than ten years, Indiana University in the USA (IU), Lund University in Sweden (LU) and Technical University of Denmark (DTU).

A High Performance Research Desktop is the concept of making high performance computing and storage systems available to users through a desktop environment. It is High Performance because it leverages High Performance Computing (HPC) compute and storage systems, and it is intended for research use as opposed to administrative or office use. For reading convenience, High Performance Research Desktops are referred to as HPC Desktops in this paper. HPC Desktops are typically deployed alongside HPC Systems, as a gateway to one or more HPC systems. However, some implementations of an HPC Desktop are not just a gateway, but also function as an environment to perform moderately parallel computational work, especially through graphical applications like MATLAB, R-Studio or similar types of applications. Users can also conveniently run visualization applications like VMD or ParaView on an HPC Desktop, without having to copy data sets or connect to remote servers. It is the goal of HPC Desktops to offer an environment that is more convenient and faster than a user's laptop or workstation. However, there will always be tasks where an HPC Desktop is not as convenient or as performant as a user's local computer~\cite{IURTIsHPCFaster}.

HPC Desktops are deployed by a number of HPC centers and research computing organizations. Among them are the organizations represented in this paper, but also others like Purdue University~\cite{Purdue}, National Laboratory for High Performance Computing Chile~\cite{NLHPC}, and the University of Chicago~\cite{UChicago}. IU and Purdue's HPC Desktop systems were described in~\cite{Thota}. There are a number of remote access solutions that are used to implement HPC Desktops, such as ThinLinc~\cite{Thinlinc}, NoMachine~\cite{Nomachine}, and FastX~\cite{Fastx}. The authors of this paper are primarily familiar with ThinLinc. The unifying concept behind all implementations of an HPC Desktop is that users have access to a persistent Linux Desktop, and can use all the built-in tools of a desktop to organize and conduct their research. Tools such as a graphical file browser, a graphical editor and access to a graphical menu that makes it easy to launch applications. Some HPC Desktop implementations go further by providing graphical tools to interact with the batch job or launch applications remotely~\cite{GfxLauncher}. In short, an HPC Desktop is more than just enabling access to remote graphical applications, it is an environment where users can implement the full workflow of their computational research.

HPC Desktops are one way to enable interactive HPC. The interactive HPC community is well established and has been holding workshops and BoFs at the International Supercomputer Conference as well as the Supercomputing Conference over the last years~\cite{InteractiveHPC}. The community has also released a state of the practice white paper recently~\cite{reuther2024interactive}. However, interactive HPC is a much broader topic than what is discussed in that paper. This paper focuses specifically on use cases that are enabled by providing a persistent, convenient and performant HPC Desktop environment.

Section \ref{CapabilitiesOfHPCDesktop} outlines the technical capabilities of an HPC Desktop. It describes hardware and software features, and also focuses on the policies that are needed to make an HPC Desktop as useful as possible. Section \ref{UseCases} is the main content of this paper, presenting a wide variety of use cases of an HPC Desktop. Section \ref{Future} briefly lists future developments for the HPC Desktops, ranging from very tangible next steps to more speculative abstraction layers. The paper ends with a conclusion and acknowledgement section.

\section{Capabilities of a High Performance Research Desktop} \label{CapabilitiesOfHPCDesktop}
This section starts with explicitly spelling out the guiding principle behind all design decisions of an HPC Desktop. This guiding principle runs counter to how HPC systems are operated, but is vital for the success of HPC Desktops. The remainder of this section outlines the architecture, policies and features of an ideal HPC Desktop. An ideal setup may not have been realized in a production environment yet, but IU, LU, and DTU operate systems that implement the majority of the features outlined below.

\subsection{The Guiding Principle of HPC Desktops}
As briefly alluded to in the introduction, the guiding principle behind the HPC Desktop is to lower the barrier of entry to HPC systems by providing users with a convenient and stable environment that performs comparably to their laptops or desktops. This principle drives all aspects of the architecture and design of an HPC Desktop. Ideally, this means that user convenience should be considered first for all hardware, software, and policy decisions. But service providers will still need to balance operational needs, hardware availability, and demand. Moreover, these HPC Desktops are a shared resource, with multiple users sharing the same compute, storage, and software environment, which adds constraints to the usage policies. This is in stark contrast to how normal HPC systems are operated, where efficient allocation of hardware and delivering the best computational and I/O performance are considered first. By providing an HPC Desktop along side an HPC system, users are provided an environment to perform light computational and data management work outside of the HPC system, leaving the HPC system open for computationally efficient workflows.

 \subsection{HPC Desktops and Open OnDemand}
Open OnDemand (OOD) is a widely used open-source software used to deploy web and graphical interfaces for applications~\cite{OOD-users,OOD}. An OOD deployment tackles many of the same use cases and scenarios that HPC Desktops address, but there are key differences. OOD also supports access to HPC storage resources in a GUI environment and other data management functions, but for computational needs, users are expected to submit jobs and launch applications on compute nodes via the OOD interface. This, in our opinion, is the major difference: while HPC Desktops can support light computational workloads as well as job submission to compute nodes, OOD only supports computational workloads on compute nodes. Using OOD, it is not possible to create a central environment where multiple applications can run on the same desktop and that users can come back to over weeks. There are pros and cons to both approaches. While it can be said that letting users do light computational work on HPC Desktops without having to submit jobs is a convenience, it also makes the service providers responsible for managing user demand and resource stability without tools designed for this purpose, like a job scheduler.

\subsection{Hardware Configuration}
\begin{figure}
\includegraphics[width=\textwidth]{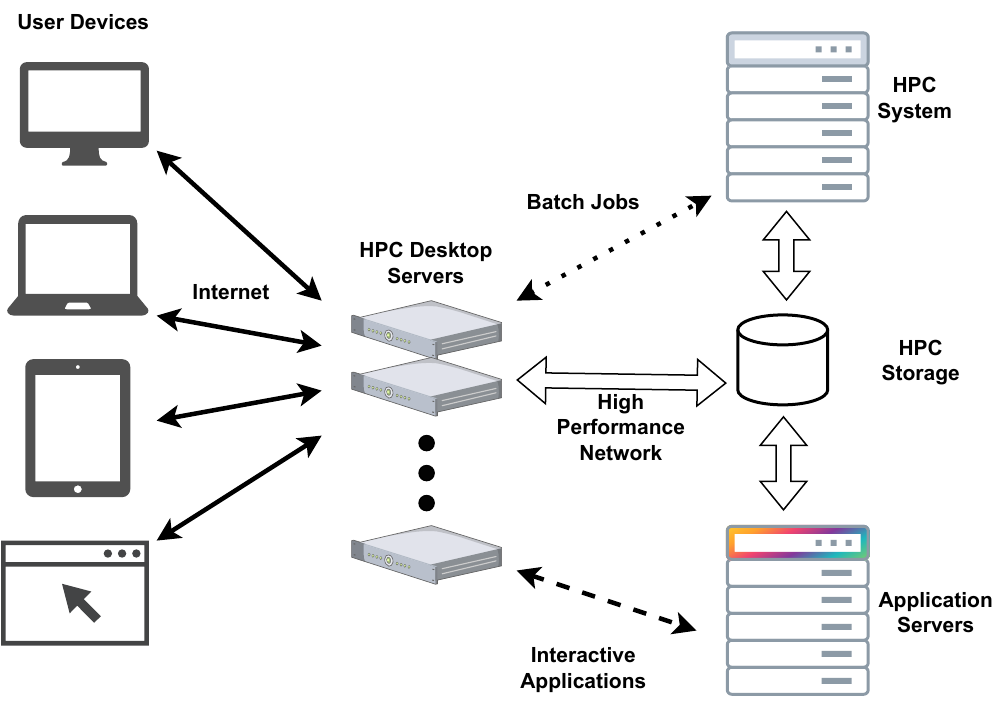}
\caption{Location of the HPC Desktop relative to High Performance compute and storage systems.} \label{HPCDesktopServers}
\end{figure}

The hardware of an HPC Desktop looks very similar to the hardware of HPC login nodes. Figure \ref{HPCDesktopServers} shows the placement of an HPC Desktop relative to an HPC system and HPC storage infrastructure. The servers that comprise an HPC Desktop can be HPC compute nodes of the current or previous generation HPC system, with a dual CPU setup and 4 to 8 GB of memory per CPU core. In following the guiding principle, enough servers need to be available to handle the expected number of concurrently active users.

IU's HPC Desktop can easily handle 15 to 20 concurrent users per server, without employing application servers. Consequently, if an average of 200 active users should be supported, about 10 servers are needed for such an environment. At LU and DTU, application servers provide dedicated capacity to run computationally demanding graphical applications. This reduces the resource requirements on the desktop servers and allows for many more users per server. Those application servers may also be used for visualization applications that require dedicated GPUs, while some HPC Desktop setups handle visualization applications by running them in the GPU partition of the HPC system via the batch system. In either case, those applications can be launched from a tool like GFXLauncher. On the network, an HPC Desktop is placed in the same location as the login nodes of an HPC system and has plenty of bandwidth to HPC file systems as well as the research network and the internet. 

\subsection{Software Setup}
The software setup of an HPC Desktop is similar to the software stack of an HPC login node. Ideally the software stack of the HPC Desktop is so similar to the software stack of HPC compute nodes that applications compiled for the HPC system can also run on the HPC Desktop, and vice versa. HPC Desktops should be able to submit jobs to the HPC system and they should mount the same file systems as HPC compute nodes and HPC login nodes so that users can work on data without having to perform data movement first.
HPC Desktop servers need to have the full X11 stack installed and depending on the remote access software that is used, additional software packages, like the Xvnc server, are needed.

\subsection{Customizing the User Desktop} \label{CustomizingDesktop}
There is a wide variety of desktop environments available, and different organizations have built their HPC Desktops on different desktop environments. Popular choices are MATE~\cite{Mate} and XFCE~\cite{XFCE}. They have been chosen for their low resource requirements, compared to modern GNOME or KDE environments~\cite{ResourceUsage}. Among the more modern desktop environments GNOME Classic~\cite{GnomeClassic} has proven to be functioning well with decent resource usage. While the desktop environment can be different, it is key to customize the actual user desktop for use as an HPC Desktop. Popular customizations include modifying the application menu of the desktop environment to include shortcuts to popular applications. Another customization is to place specific icons in the “panel” and on the desktop, icons that offer shortcuts to a terminal, the file manager or help/documentation.
Some institutions have developed utilities that help users with their first steps into batch system usage. Such tools are also commonly made available through desktop icons. GfxLauncher is one such tool. It is an open-source Qt-based Python application that can launch different interactive applications through the batch system using a configurable user interface. Currently, it supports launching X11-based applications, OpenGL accelerated applications (VirtualGL~\cite{VirtualGL}), and Jupyter notebooks. Providing a custom desktop background in line with the organizations branding is also recommended.

\subsection{Policies}
Usage policies set the HPC Desktop apart from normal HPC login nodes. On an HPC Desktop, users are encouraged to perform work that may run for a long period of time, hours or days, as long as this work does not monopolize the node. There is no hog-watch setup on an HPC Desktop and CGroups is configured to prevent users from running the node out of memory and enable fair use of a shared environment.

On the IU HPC Desktop it was observed that that users occasionally seem to abuse the nodes by running computational workloads not appropriate for a shared environment. However, most of the time users were not aware of the parallelism of the application that they were using. Over the last years, this occurred infrequently enough that it can be dealt with on a case by cases basis by educating the user. Monitoring scripts are in place to detect such events and notify support personnel.

HPC Desktop users can disconnect from and reconnect to their desktop session in a similar way to how screen works in an ssh environment. This feature allows users to come back to an environment for days and weeks. While it is tempting to reduce resource consumption by terminating sessions that no user is currently connected to, it is key for a convenient environment to allow users to reconnect to their sessions multiple days after they have disconnected. At IU and LU, reconnecting to a session is possible for up to 7 days. At DTU, the lifetime of a desktop session is not limited, but limits are enforced applications started from the desktop via GFXLauncher.

\section{Use Cases} \label{UseCases}
The use cases of this section are derived from HPC Desktops operated by IU, LU, and DTU over more than 10 years. These Desktops have attracted users from all fields of science. The use cases have been reconstructed from user interviews and from observing user support tickets.
On a very high level, the use cases can be grouped into using standard desktop features to enable research workflows and using specific HPC features like the batch system. While the second ones are more interesting, the first ones should not be underestimated because they enable a convenient user environment.

\subsection{Using a Graphical File Manager}

The ability to use a graphical file manager is a real game changer for new users in a Linux and HPC environment. While it is true that managing files on the command line only requires learning a few commands, in practice this puts a huge burden on new users. New users usually have only ever used a graphical file manager like the MacOS Finder or the Windows File Explorer, and the concept of "running a command to perform file actions" is foreign to them.
Use cases enabled by a graphical file manager include:
\begin{itemize}
    \item Easy unpacking of archive files by selecting a file with the mouse and selecting "Extract here" from the file's context menu.
    \item Create an archive from a directory tree by selecting "Compress" from the directory's context menu.
    \item Looking up how much storage is consumed by an entire directory tree by looking at directory properties or using a disk usage analyzer tool.~\cite{GNOMEDiskUsageAnalyzer}
    \item Moving files from one storage location to another by using "Copy and Paste" or "Drag and Drop" (for example from HOME to SCRATCH).
    \item File deletion, through either moving file(s) to the "Trash", or right-clicking to choose "Delete" from a pop-up menu. Having a "Trash" facility that makes it easy for users to undelete a file. This however also adds to the complexity for users to understand what files are accounted towards a user's file system quota, as large files in "Trash" can contribute to the quota.
    \item Bookmarking directories in the file manager allows users to easily remember the mount points for different file system, for example HOME, SCRATCH, PROJECT etc. 
    \item Easily connect to "outside" storage location by using the "Connect to Server" functionality. This allows for easy data movement from Windows file shares as well as connecting to cloud storage providers.
\end{itemize}

\subsection{Pre and Post Processing for HPC Jobs, Running HPC Jobs}

This section lists tasks that benefit from a graphical desktop environment when preparing, submitting and managing HPC jobs. This section also lists tasks that users perform to prepare for running software on an HPC system, for example moving data sets or installing custom software packages.

\begin{itemize}
    \item Easy movement of files between a user's machine and the HPC Desktop. ThinLinc provides a convenient thindrives~\cite{ThinDrives} mechanism for this task.
    \item Download software and data sets straight to an HPC file system using a web-browser on the HPC Desktop, especially for downloads that require web authentication.
    \item Long running data movement operations like rsync or moving files to a tape archive.
    \item Using GfxLauncher with interactive applications, such as post processors, visualization packages and notebooks. By scheduling valuable high performance graphic nodes using the batch system with a user defined wall-time, better resource utilization is achieved. This enables users to run heavy interactive applications directly on the desktop without overloading the HPC Desktop servers. At LU, this has enabled new user groups to take advantage of HPC such as humanities (language models and 3D reconstruction in archaeology). 
    \item Using a graphical debugger from an interactive job with X-Forwarding to the HPC Desktop.
    \item Using a graphical IDE to write, test, and debug code in an environment that mimics the destination HPC environment.
    \item Running graphical HPC performance analysis tools like Vampir, MAP, the gprofng GUI~\cite{gprofngGUI}, or NVIDIA Nsight.
    \item Graphical editors to create and edit HPC batch scripts.
    \item Quickly look at visualizations created by HPC jobs using an image viewer.
\end{itemize}

\subsection{Non HPC Work}

The HPC Desktop provides a convenient environment for running graphical applications for a long time, especially if those applications are serial or modestly parallel so that they run well in a shared environment. These applications can access data on large central file systems without having to copy the data. The ability of the user to disconnect and reconnect to an HPC Desktop over the course of days and weeks makes it easy to run an application for a long time, something that may be difficult to accomplish on a laptop. The HPC Desktop allows for sharing access to licensed software packages with a limited number of licenses. Depending on the license model, users can share the software on a central system without having to install and license the software. Some of the non-HPC tasks enabled by an HPC Desktop are:
\begin{itemize}
    \item Long running Jupyter Notebooks or R-Studio sessions.
    \item Access to statistical software like SPSS, SAS, STATA (support long running stats scripts that are not parallel, but need to run for a long period of time).
    \item Run applications that are only licensed on central system. (shared license model for MATLAB, SPSS or Photoscan or other vended software packages)
    \item Visualization Software, with local or remote use of GPUs using VirtualGL.
    \item Run applications that operate on large data but do not require intensive CPU or excessive memory usage.
\end{itemize}

\subsection{Teaching and Learning}

An HPC Desktop can facilitate teaching and learning, not just for classes that require HPC systems. The obvious use case is that HPC centers and research computing organizations offer "Introduction to HPC" classes using the HPC Desktop. Students that take the class can continue to experiment and learn after class, by reconnecting to their HPC Desktop, potentially with folders and terminals still open on the desktop. This makes it much more likely that students continue using the system compared to when they connect with SSH and are presented with a "new shell" every time they connect.

Another use case is faculty members leveraging an HPC desktop in their class. Students can connect to the desktop and are presented with a pre-configured environment ready to use. This is especially relevant if the HPC Desktop offers the ability to connect via a web-based user interface. This eliminates the need for students to install an application and allows student with lower-end devices to participate equally. Another advantage of leveraging a centrally provided system is that students can get access to different software packages, for example different FORTRAN compilers or engineering applications, without having to install that software locally.

\subsection{Client Server Applications}
There are a number of research applications that utilize a client server compute model, where usually a graphical client applications spawns a server application that handles compute intensive tasks. Example applications that have this capability include, MATLAB, COMSOL Multiphysics, ANSYS Workbench, ABAQUS CAE, and Schrödinger. The graphical client of these applications can run on an HPC Desktop and the server can run in the batch system. Since HPC Desktops are located next to the HPC system, the client can directly communicate with the server, without having to route that communication through an SSH tunnel. This makes it a lot easier for users to utilize such client server applications, compared to running the client on a laptop and then having to manually establish an SSH tunnel to the HPC system to enable communication between client and server. Specific use cases include:

\begin{itemize}
    \item At IU, the graphical versions of COMSOL Multiphysics and Schrödinger are available on the HPC Desktop. They can be used to perform small computations the HPC Desktop and users can send jobs to the cluster through the GUI and review results in the interface when jobs are completed. In IU's case, setting this up took some staff time to properly design and maintain the configuration files (one for each cluster to which the user is submitting jobs).
    \item At DTU and LU, MATLAB Parallel Server is tightly integrated into the MATLAB Desktop, that can be started from the HPC Desktop session.  Via different cluster profiles, MATLAB workers can either run locally, or are dispatched to the HPC nodes.
    \item At IU, there are a number of users who use Jupyter Notebook or Jupyter Lab. In some cases, they need to use GPUs that are not available in IU's HPC Desktop, or they need more compute or memory resources than the shared environment of an HPC Desktop provides. Launching those notebooks in the batch system and running the web browser on the HPC Desktop is straightforward. Alternatively, X11 forwarding can be used to display a web browser running on a compute node on the HPC Desktop.
\end{itemize}

\subsection{Secure Enclaves}
Secure enclaves enable users to work with sensitive data sets, for example electronically protected health information, restricted research data or licensed third-party data sets with restrictive data use agreements. HPC Desktops can dramatically increase the usability of secure enclaves by providing a graphical desktop with most of the features outlined above. The specific implementation of a secure enclave depends heavily on the sensitivity of the data and the standards that the enclave needs to conform to, for example NIST 800-53/171 or FISMA. IU has used an HPC Desktop to provide researchers from multiple universities access to a secure enclave to utilize graphical statistical applications like SPSS, SAS, or RStudio.

\section{Future Developments} \label{Future}

This section speculates on future developments for the HPC Desktop. Some of the items outlined are very technical and are logical next steps to keep up with the development of technology and the evolution of Linux desktop environments. Other items are more speculative and represent a real evolution of the concept of making an HPC system accessible via a desktop metaphor.

\subsection{Experiment with State of the Art Desktop Environments}

Today's HPC Desktops use older or very light weight desktop environments like MATE or XFCE. The reason is that these desktops consume fewer resources, which is an important consideration for a shared environment with dozens of users per machine. However, just as the Windows and MacOS desktops have evolved over the last years, so have Linux desktop environments. Providing an HPC Desktop with Gnome 45 or the latest KDE Plasma may provide users with a more compelling environment.

\subsection{Explore deep Integration of an HPC System into the Desktop}

As outlined in section \ref{CustomizingDesktop}, an HPC Desktop features a customized desktop that exposes some of the features of an HPC system in a user friendly way. Modern desktop environments and file browsers make it easy to develop plugins or extensions that can take customizations to the next level. A few things to explore are: \vspace{-7mm}

\subsubsection{Run In Batch}

When selecting the context menu of an application, offer the option to launch this application in the batch system rather than on the desktop. This can be accomplished in a number of different ways, for example by using GfxLauncher with the correct parameters or by running the application inside an interactive job. By integrating this option into the desktop environment, it becomes available for every application rather than having to provide separate icons for running the application on the desktop and in the batch system. \vspace{-7mm}

\subsubsection{Recently Run Jobs}

File browsers feature a Recently Opened Files section. With a plugin, the file browsers could feature a Recently Run Jobs section, that would provide a list of JobIDs and JobNames. When a job is selected, the file browser would show the location of the job script. \vspace{-7mm}

\subsubsection{Move to Archive}

Institutions with an HPC Desktop are very likely to also provide a long term archive for their users. Enhancing the file browser with the ability to archive a directory tree would significantly enhance the usability of such an archive and could also enforce a minimum set of metadata to be associated with the data. For example a user could select a directory to be copied to the archive and a dialog would open prompting the user for metadata, before compressing the directory tree into an archive file and moving it to the archive.

\subsection{Graphical Tools for HPC Tasks}

One barrier to entry in HPC is the many command line tools required to accomplish common HPC operations such as loading software modules, creating job scripts and managing running jobs. Providing a research desktop environment provides an environment where it is easy to create simple graphical user interfaces for these tasks. LUNARC has developed an LMOD browser tool, ml-browse~\cite{mlbrowse}, that make it easy find available software using a searchable browser interface. The GfxLauncher framework also provides a graphical job manager for monitoring and controlling running jobs. Work is also ongoing to develop a wizard for creating SLURM job scripts for common tasks. 

For the Archeology department at LU a workflow was developed for running 3D reconstruction using the Metshape software, which is a graphical application with it's own batch framework. To make this workflow even easier, a graphical tool will be provided in the future to bootstrap the batch framework.

\subsection{Develop a Community around the HPC Desktop}

While institutions have made their interactive tools available and have documented them on their support pages or on Github, there is no central location where "all things HPC Desktop" can be found. If such a place were to be created, it could help with:
\begin{itemize}
  \item Share scripts and desktop customizations,
  \item Share experiences of what worked and didn't work when engaging users,
  \item Share statistics scripts and ideas for metrics.
\end{itemize}

\section{Conclusion}
HPC Desktops have proven successful in attracting new users to HPC systems and have broadened the user base for many institutions, including Indiana University, Lund University and Technical University of Denmark. HPC Desktops have enabled use cases that would be hard or impossible to support with traditional "ssh-only" HPC systems. 
HPC Desktop's unique contribution to the HPC ecosystem is to provide users with a single environment in which to perform all their computation research. This sets them apart from solutions like domain-specific science gateways or web based HPC access methods like Open OnDemand.
HPC Desktops are in use by many HPC centers and research computing organizations around the world, but they are not yet a mainstream tool. Fostering a vibrant community around HPC Desktops will be instrumental in sharing best practices, innovations, and experiences, and ensuring that HPC Desktops continue to serve as an enabler of scientific discovery and innovation. 

\section{Acknowledgements}
We would like to acknowledge the organizers of the Interactive and Urgent HPC BoFs and workshops at ISC and SC. We hope to leverage these events in the future to build a community around HPC Desktops. We would also like to acknowledge the many HPC centers and research computing organizations that provide an HPC Desktop or similar tool for their users.

%
%
%

\begin{thebibliography}{8}

\bibitem{IURTIsHPCFaster}
Determine whether high performance computing can help your research, \url{https://kb.iu.edu/d/bilm}.

\bibitem{Purdue}
RCAC - Knowledge Base: Scholar User Guide: ThinLinc, \url{https://www.rcac.purdue.edu/knowledge/scholar/accounts/login/thinlinc}.

\bibitem{NLHPC}
NLHPC Laboratorio Nacional de Computacion de Alto Rendimiento, \url{https://www.nlhpc.cl}

\bibitem{UChicago}
University Of Chicago Basics - RCC User Guide, \url{https://rcc-uchicago.github.io/user-guide/thinlinc/}

\bibitem{Thota}
A. Thota, , D. Dietz, C. Phillips, X. Zhu, L. Weakley, B. Fulton, H. Dennis, L. Huber, S. Michael, W. Snapp-Childs, S. Harrell, A. Younts: Research Computing Desktops: Demystifying research computing for non-Linux users. 2019. \url{https://dl.acm.org/doi/10.1145/3332186.3332206}

\bibitem{Thinlinc}
Linux Remote Desktop based on open-source - ThinLinc by Cendio, \url{https://www.cendio.com}

\bibitem{Nomachine}
NoMachine, \url{https://www.nomachine.com}

\bibitem{Fastx}
FastX - Starnet, \url{https://www.starnet.com/fastx}

\bibitem{GfxLauncher}
GfxLauncher, \url{https://gfxlauncher-documentation.readthedocs.io/en/latest/}

\bibitem{InteractiveHPC}
InteractiveHPC web page, \url{https://www.interactivehpc.com}

\bibitem{reuther2024interactive}
A. Reuther, N. Brown, W. Arndt, J. Blaschke, C. Boehme, A. Chazapis, B. Enders, R. Henschel, J. Kunkel, M. Martinasso: Interactive and Urgent HPC: Challenges and Opportunities. 2024. \url{https://arxiv.org/abs/2401.14550}

\bibitem{OOD-users}
OpenOnDemand Case Studies, \url{https://openondemand.org/get-involved\#case-studies}

\bibitem{OOD}
Open OnDemand: Transforming Computational Science Through Omnidisciplinary Software Cyberinfrastructure, David E. Hudak, Douglas Johnson, Jeremy Nicklas, Eric Franz, Brian McMichael, Basil Gohar, Proceedings of the XSEDE16 Conference on Diversity, Big Data, and Science at Scale, Miami, USA, 2016. 

\bibitem{Mate}
MATE Desktop Environment, \url{https://mate-desktop.org}

\bibitem{XFCE}
XFCE Desktop Environment, \url{https://www.xfce.org}

\bibitem{ResourceUsage}
Desktop Environments Resource Usage Comparison, \\\url{https://vermaden.wordpress.com/2022/07/12/desktop-environments-resource-usage-comparison/}

\bibitem{GnomeClassic}
What is GNOME Classic,\\
\url{https://help.gnome.org/users/gnome-help/stable/gnome-classic.html}

\bibitem{VirtualGL}
VirtualGL, \url{https://www.virtualgl.org/}

\bibitem{GNOMEDiskUsageAnalyzer}
GNOME Disk Usage Analyzer, \\
\url{https://community.linuxmint.com/software/view/baobab}

\bibitem{ThinDrives}
Accessing Local Drives with ThinDrives, \\
\url{https://www.cendio.com/resources/docs/tag/redir\_drives.html}

\bibitem{gprofngGUI}
gprofng GUI, \url{http://savannah.gnu.org/projects/gprofng-gui/}

\bibitem{mlbrowse}
LMOD Graphical module browser,
\url{https://github.com/lunarc/mbrowser}


\end{thebibliography}
%

\end{document}